\documentclass[fleqn,twoside]{article}
\usepackage{espcrc2}

\usepackage{graphicx}

\newcommand{\AmS}{{\protect\the\textfont2
  A\kern-.1667em\lower.5ex\hbox{M}\kern-.125emS}}

\title{
\vspace{-4.2cm}
\begin{flushright}
{\normalsize
hep-lat/0110143\\
October 2001}
\end{flushright}
\vspace{2.2cm}
Hadron Spectrum for Quenched Domain-Wall Fermions with DBW2 Gauge
Action\thanks{Talk presented at Lattice 2001, Berlin, Germany.}
}

\author{Yasumichi Aoki\address{RIKEN BNL Research Center, Brookhaven
National Laboratory, Upton, NY 11973, USA} [RBC collaboration]
\thanks{We thank RIKEN, Brookhaven National Laboratory and the U.S.\
Department of Energy for providing the facilities essential for the
completion of this work.}}

\begin{document}

\begin{abstract}
We investigate basic physical quantities for quenched simulation
with domain-wall fermions and the DBW2 gauge action. Masses
and decay constant of pseudoscalar mesons
are measured. Scaling properties are tested.
\end{abstract}

\maketitle

\section{INTRODUCTION}

Domain wall fermion (DWF)\cite{Kaplan,FurmanShamir} is a promising tool
to calculate hadronic quantities 
where chiral symmetry is important. Precise calculation of kaon
mixing parameter $B_K$ is one success of this
formulation\cite{BK}. More challenging 
calculations\cite{KpCPPACS,KpRBC} of kaon weak matrix element for
$\epsilon'/\epsilon$ 
show a first reliable determination within lowest order chiral
perturbation theory. 
Attempt to calculate nucleon matrix elements is also underway\cite{GAlat}. 

The good chiral property of DWF comes at an obvious cost, 
which is a substantial increase of computational effort associated 
with the size of the fifth dimension $L_s$. This makes taking the
continuum limit by performing lattice finer than $a^{-1}\simeq 2$ GeV
difficult. 
Going to a coarser lattice  to have another point for the continuum
extrapolation could be a better strategy.
However, results\cite{DWCPPACS,DWRBC} on coarser lattices show that 
an incredibly large size for the fifth dimension is required to keep the
residual chiral symmetry breaking small. 

Motivated by the fact that improving the gauge action reduces the 
residual chiral symmetry breaking at finite $L_s$ significantly
at $a^{-1}\simeq 2$ GeV\cite{DWCPPACS}, we found the DBW2 action\cite{DBW2}, 
which is a non-perturbatively determined Iwasaki-type
action\cite{Iwasaki}, gave much more improvement for the chiral
properties\cite{KostasLat01,TakuLat01}. 
Here we investigate basic physical quantities for the DBW2 gauge action for
quenched DWF to see the
viability of this approach down to $a^{-1}\simeq 1.3$ GeV. 

We perform the simulation on a $16^3\times 32$ lattice with DBW2 action
at $\beta=0.87$ and $1.04$ in Iwasaki type 
parameterization, which corresponds to $a^{-1}\simeq 1.3$ and $2$ GeV
respectively. 
The domain wall hight $M_5$ is tuned in the first decimal place so that
it minimizes  
the residual quark mass $m_{res}$, although it is not necessary that 
it be done to this accuracy. 

\begin{table}[t]
 \caption{Simulation parameters and calibration.}
 \label{tab:param}
 \begin{tabular}{cccccc}
  \hline
  $\beta$ & $M_5$ & $L_s$ & stat. & \mbox{\hspace{-4pt}}$m_{res}a$ & \mbox{\hspace{-6pt}}$a_\rho^{-1} \mbox{(GeV)}$ \\
  \hline
  0.87  & 1.8 & 16 & 100 & $5.7(3) 10^{-4}$  & $1.28(3)$\\
  1.04  & 1.7 & 16 & 200 & $1.7(1) 10^{-5}$  & $1.97(4)$\\
  \hline
 \end{tabular}
\vspace{-20pt}
\end{table}

\section{RESULTS FOR $a^{-1}\simeq 1.3$ GeV}

The residual quark mass $m_{res}$ is defined through the ratio
\begin{equation}
 R(t) = \left<J_{5q}(t)P(t)\right>/\left<P(t)P(0)\right>
\end{equation}
at $t\gg a$,
where $J_{5q}$ represents explicit breaking of chiral symmetry 
in the the divergence of ``conserved current'' $\cal A_\mu$,
and $P$ is the pseudoscalar density operator.
We list the value of $m_{res}$ in Table \ref{tab:param}.
For $a^{-1}\simeq 1.3$ GeV unrenormalized value of $m_{res}\simeq 0.7$ MeV,
which is smaller than that obtained for the Wilson gauge action at
finer lattice, $a^{-1}\simeq 2$ GeV and the same $L_s$,
$m_{res}\simeq 2.4$ MeV\cite{DWRBC}.

\begin{figure}[t]
 \begin{center}
  \includegraphics[width=7cm]{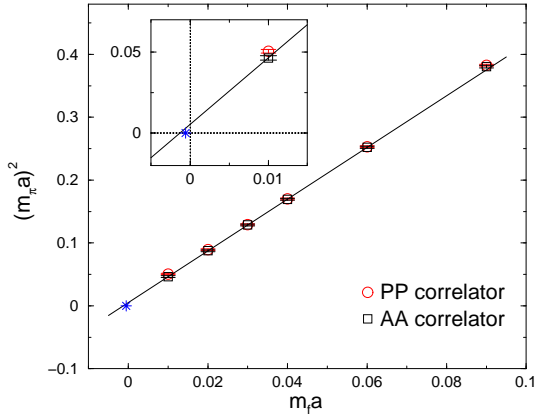}
 \end{center}
\vspace{-32pt}
 \caption{Pion mass squared as a function of quark mass for
 $a^{-1}\simeq 1.3$ GeV. Asterisk indicates the point $m_f=-m_{res}$.}
\label{fig:pi}
\vspace{-20pt}
\end{figure}

The pion mass can be extracted from a two point function constructed from
pion interpolating operators.
We consider two different two-point
functions. One is the pseudoscalar -- pseudoscalar, $\left<PP\right>$,
the other is the local axial -- axial correlator,
$\left<A_0 A_0\right>$. The former has a $1/m^2$ and $1/m$ pole from zero modes
in finite volume, while the later has only $1/m$. 
Figure \ref{fig:pi} shows the pion masses from those different two
point functions 
are consistent with each other down to $m_f=0.02$. The difference begins to 
show up at the lightest mass $m_f a = 0.01$, however it is quite small
compared to the value of mass itself. 

A linear fit of the  $\left<A_0 A_0\right>$ correlator for $0.01\le m_f \le
 0.06$ is quite good ($\chi^2/dof=0.03/3$). However, x-intercept is 
 $m_f=-1.3(4)  10^{-3}$, so it overshoots  $m_f=-m_{res}$,
where the pion mass should vanish. This location of the intercept is guaranteed
with accuracy of $O(a^2)$ by the Ward-Takahashi identity when
all possible finite size effects have been removed\cite{DWRBC}.
With this constraint one can fit to a quenched chiral log formulae,
\begin{eqnarray}
 (m_{\pi}a)^2 & = & a_0 m \left(1 + a_1 \log m\right) + a_2 m^2\\
 m & \equiv & m_f a + m_{res} a.
 \end{eqnarray}
Results are listed in the Table \ref{tab:fit}.
Both fits with or without quadratic term give reasonable value of $a_1$.

\begin{table}[b]
\vspace{-16pt}
\caption{Fit result of pion mass. ``-'' indicates that the parameter is
 constrained to be zero.}
\label{tab:fit}
\begin{tabular}{ccccc}
 \hline
 $m_f$ & $a_0$ & $a_1$ & $a_2$ & $\chi^2/dof$ \\
 \hline
 0.01-0.06 &  3.9(2) & 0.03(2) & - & 0.5/3\\
 0.01-0.09 &  3.1(5) & 0.09(5) & 4.9(23) & 0.04/3\\
 \hline
\end{tabular} 
\end{table}

\begin{figure}[t]
 \begin{center}
 \includegraphics[width=7cm]{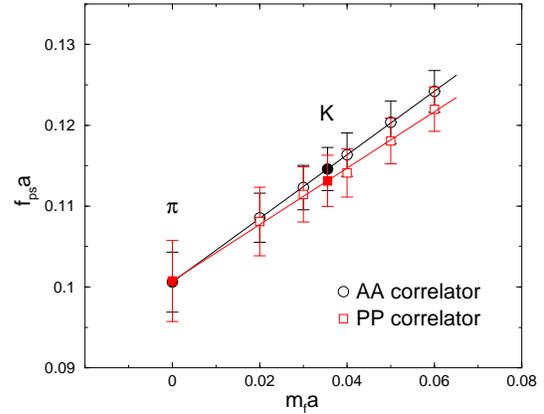}
 \end{center}
\vspace{-32pt}
 \caption{Pseudoscalar decay constant as a function of quark mass at
 $a^{-1}\simeq 1.3$ GeV.}
\label{fig:fps}
\vspace{-20pt}
\end{figure}

We determine the pseudoscalar decay constant using two methods\cite{DWRBC}.
One is an estimate from the $\left<A_0 A_0\right>$ correlator
with the renormalization factor $Z_A$
nonperturbatively determined from the ratio of $\left<{\cal
A}_0 P\right>$ and $\left<A_0 P\right>$. The other uses 
the $\left<P P\right>$ correlator with the equation $\Delta_\mu {\cal A_\mu}
\simeq 2(m_f+m_{res})P$ from the Ward-Takahashi 
identity. As seen in the Figure \ref{fig:fps} the two methods give
consistent results, 
indicating the effect of chiral symmetry breaking can be absorbed in
the shift of $m_f$.
Also the possible zero mode effects, which influence the two 
correlators differently, appear to be small, 
which is a confirmation of what we saw for the pion mass down
to $m_f=0.02$.

\section{SCALING PROPERTIES}

\begin{figure}[t]
 \begin{center}
 \includegraphics[width=7cm]{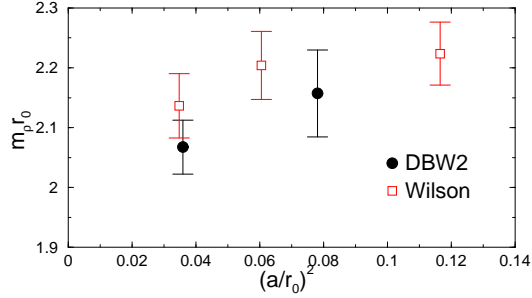}
 \end{center}
\vspace{-36pt}
 \caption{Consistency check of scales from the rho meson mass and
 the heavy quark potential.}
\label{fig:rho}
\vspace{-20pt}
\end{figure}

Figure \ref{fig:rho} plots the rho meson mass extrapolated to the chiral limit as a
function of lattice spacing. 
Both axes are normalized by $r_0$ calculated from the heavy quark
potential. Together with the data of this work, results with
Wilson gauge action\cite{DWRBC} are shown. Those data show that the scales
determined by different methods agree very well for both the DBW2 and
the Wilson gauge action for DWF.
This good scaling is also seen in the nucleon mass and $K^{*}$ mass
normalized by rho meson mass. This also holds when one chooses $r_0$ for
the input of scale, which is obvious from above discussion.

\begin{figure}[b]
\vspace{-18pt}
 \begin{center}
 \includegraphics[width=7cm]{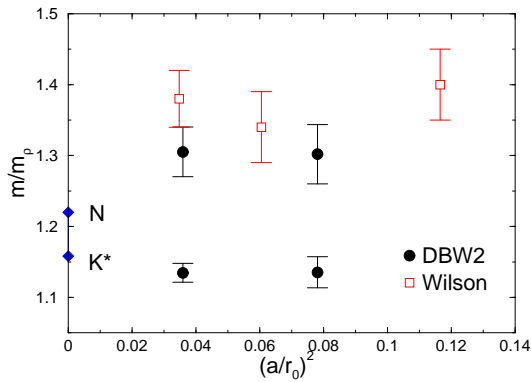}
 \end{center}
\vspace{-36pt}
 \caption{Scaling plot of nucleon and $K^*$ mass. Diamonds are
 experimental values.}
\end{figure}

\begin{figure}[h]
 \begin{center}
 \includegraphics[width=7cm]{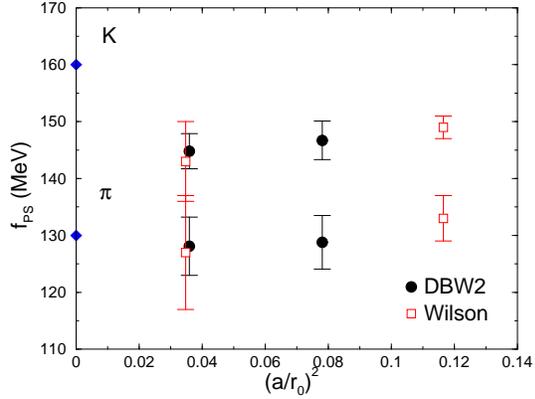}
 \end{center}
\vspace{-36pt}
 \caption{Scaling plot of pseudoscalar decay constant. Diamonds are
 experimental values.}
\label{fig:fpsa}
\vspace{-20pt}
\end{figure}

Scaling of the pseudoscalar decay constant (Fig.~\ref{fig:fpsa}) is also good.
$f_{\pi}$ 
is in good agreement with the experimental value. $f_K$
appears to be smaller than the experimental value, which could
be caused by the effects of quenching\cite{BernardGolterman}.

\section{CONCLUSION}
We have attempted to use the DBW2 gauge action in a quenched domain wall fermion
simulation. Good chiral properties for $a^{-1}\simeq 1.3$ GeV has been
observed. We have investigated scaling of hadronic observables, which also turned
out to be good. We have also confirmed that the values were
consistent with those obtained with the Wilson gauge action.

The statistics presented here is not sufficient for the precision measurement
for those observables calculated here. However, we can argue that 
there is no evidence for any difficulty with this choice of gauge
action, 
rather it has lots of good features.

An important benefit from the use of this gauge action is that residual
chiral symmetry 
breaking is quite small even for a coarse lattice with $a^{-1}\simeq 1.3$ GeV. 
Therefore it may be suitable for 
demanding calculation like $\epsilon'/\epsilon$.
Once we have the results for $a^{-1}\simeq 1.3$ GeV and finer lattices, we
will be able to discuss the scaling property for those quantities.

\end{document}